\documentstyle[12pt]{article}
\newcommand\be{\begin{equation}}
\newcommand\ee{\end{equation}}
\newcommand\bea{\begin{eqnarray}}
\newcommand\eea{\end{eqnarray}}
\newcommand\ket[1]{|#1\rangle}

\newcommand\braket[2]{\langle #1|#2\rangle}
\newcommand{\fatalpha}{{\bf \alpha \kern -0.44em \alpha}}
\newcommand{\fatsigma}{{\bf \sigma \kern -0.54em \sigma}}
\newcommand{\tpchi}{{\bf \chi \kern -0.35em \chi}}
\newcommand{\llambda}{{\bf \lambda \kern -0.45em \lambda}}

\renewcommand{\theequation}{\arabic{equation}}
\renewcommand{\theequation}{\thesection-\arabic{equation}}
\bibliography{plain}
\title{\bf Investigation of continuous-time quantum walks via spectral analysis and Laplace transform}\vspace{20mm}
\author{ M. A. Jafarizadeh$^{a,b,c}$
 \thanks{E-mail:jafarizadeh@tabrizu.ac.ir},
 R. Sufiani$^{a,b}$
 \thanks{E-mail:sofiani@tabrizu.ac.ir}
\\ $^a${\small Department of Theoretical Physics and Astrophysics,
University of Tabriz, Tabriz 51664, Iran.} \\ $^b${\small
Institute for Studies in Theoretical Physics and Mathematics,
Tehran 19395-1795, Iran.} \\ $^c${\small Research Institute for
Fundamental Sciences, Tabriz 51664, Iran.}} \pagebreak

\vspace{20mm}
\begin{document}
\maketitle \vspace{15mm}
\newpage
\begin{abstract}
Continuous-time quantum walk (CTQW) on a given graph is
investigated by using the techniques of the spectral analysis and
inverse Laplace transform of the Stieltjes function (Stieltjes
transform of the spectral distribution) associated with the graph.
It is shown that, the probability amplitude of observing the CTQW
at a given site at time $t$ is related to the inverse Laplace
transformation of the Stieltjes function, namely, one can
calculate the probability amplitudes only by taking the inverse
laplace transform of the function $iG_{\mu}(is)$, where
$G_{\mu}(x)$ is the Stieltjes function of the graph. The
preference of this procedure is that, there is no any need to know
the spectrum of the graph.
\\


{\bf PACs Index: 03.65.Ud }
\end{abstract}

\vspace{70mm}
\newpage
\section{Introduction}
One of the most challenging problems in quantum computation has been
the design of quantum algorithms which outperform their classical
counterparts in meaningful tasks. Few quantum codes in this category
have been discovered after the well-known examples by Shor and
Grover \cite{shor,grover}. A natural way to discover new quantum
algorithmic ideas is to adapt a classical one to the quantum model.
An appealing well-studied classical idea in statistics and computer
science is the method of random walks \cite{RW1,RW2}. Recently, the
quantum analogue of classical random walks has been studied in a
flurry of works \cite{5', 6', 7', 8', 9', 10'}. The works of Moore
and Russell \cite{9'} and Kempe \cite{10'} showed faster bounds on
instantaneous mixing and hitting times for discrete and continuous
quantum walks on the hypercube (compared to the classical walk). The
focus of this paper is on the continuous-time quantum walk that was
introduced by Farhi and Gutmann \cite{5'}. In Refs.
\cite{js,js1,rootlatt}, the dynamical properties of CTQW on finite
and infinite graphs have been studied by using the spectral
analysis. In fact, it has be shown that the probability amplitudes
of the CTQW on a given graph are related to the Stieltjes function
or Stieltjes transformation of the spectral distribution associated
with the adjacency matrix of the graph, i.e., in order to calculate
the probability amplitudes one needs to evaluate the inverse
Stieltjes transformation. The Stieltjes function has extensive
applications for example in solid state physics and condensed
matter, where it is known as Green function. It can be also used for
calculating the two-point resistances on regular resistor networks
\cite{ress}. Although, with any arbitrary graph one can associate a
Stieltjes function (see Ref.\cite{js2}), calculation of the inverse
Stieltjes transformation is not an easy task. In this work, we show
that the probability amplitudes of CTQW on graphs can be evaluated
by taking the inverse Laplace transformation of the Stieltjes
function which is easier and more popular transformation with
respect to the inverse Stieltjes transformation. In fact, the
Laplace transformation is used extensively in various problems of
pure and applied mathematics. Particularly widespread and effective
is its application to problems arising in the theory of operational
calculus and in its applications, embracing the most diverse
branches of science and technology. An important advantage of
methods using the Laplace transformation lies in the possibility of
compiling tables of direct and inverse Laplace transforms of various
elementary and special functions commonly encountered in
applications. We survey and re-derive equations for the CTQW on QD
\cite{js1,obata} and non-QD \cite{js2} type graphs by using the
Laplace transformation.

The organization of the paper is as follows: In section $2$, some
preliminary facts about graphs and their stratifications, the
quantum decomposition for the adjacency matrix of some particular
graphs called QD graphs and Stieltjes transform of spectral measure
associated with the graph are reviewed. In section $3$, we review
CTQW on an arbitrary graph and give a procedure for calculating the
probability amplitudes of CTQWs on graphs by using the inverse
Laplace transformation. Section $4$ is devoted to some examples of
CTQW on QD and non-QD type graphs and calculation of the
corresponding probability amplitudes of the walk on them by
employing the techniques introduced in the section $3$. The paper is
ended with a brief conclusion and an appendix containing a table of
examples of CTQW on some important distance-regular graphs.
\section{Preliminaries}
In this section we review some preliminary facts about graphs and
their stratifications, the quantum decomposition for the adjacency
matrix of some particular graphs called QD graphs and Stieltjes
transform of spectral measure associated with the graph.
\subsection{Graphs and their stratifications}
A graph is a pair $G=(V,E)$, where $V$ is a non-empty set and $E$
is a subset of $\{(i, j); i, j\in V, i\neq j\}$. Elements of $V$
and of $E$ are called vertices and edges, respectively. Two
vertices $i, j\in V$ are called adjacent if $(i, j)\in E$, and in
that case we write $i\sim j$. A finite sequence $i_0; i_1; ...;
i_n\in V$ is called a walk of length $n$ (or of $n$ steps) if
$i_{k-1}\sim i_k$ for all $k=1, 2, ..., n$. A graph is called
connected if any pair of distinct vertices is connected by a walk.
The degree or valency of a vertex $x\in V$ is defined by
$\kappa(x)=|\{y\in V: y\sim x\}|$. The graph structure is fully
represented by the adjacency matrix $A$ defined by
\begin{equation}\label{adj.}
\bigl(A)_{i,j}\;=\;\cases{1 & if $\; i\sim j$,\cr 0 &
otherwise\cr}\qquad \qquad (i, j \in V).
\end{equation}
Obviously, (i) $A$ is symmetric; (ii) an element of $A$ takes a
value in $\{0, 1\}$; (iii) a diagonal element of $A$ vanishes. Let
$l_2(V)$ denote the Hilbert space of square-summable functions on
$V$, and $\{|i\rangle ; i\in V \}$ becomes a complete orthonormal
basis of $l_2(V)$. The adjacency matrix is considered as an
operator acting in $l_2(V)$ in such a way that
\begin{equation}\label{adj1.}
A\ket{i}=\sum_{j\sim i}\ket{j},\;\;\ i\in V.
\end{equation}

For $i\neq j$ let $\partial(i, j)$ be the length of the shortest
walk connecting $i$ and $j$. By definition $\partial(i, j)=0$ for
all $i\in V$ . The graph becomes a metric space with the distance
function $\partial$. Note that $\partial(i, j)=1$ if and only if
$i\sim j$. We fix a point $o\in V$ as an origin of the graph.
Then, a natural stratification for the graph is introduced as:
\begin{equation}\label{str.}
V=\bigcup_{i=0}^{\infty}V_i(o) \;\ , \;\;\ V_i(o):=\{j\in V :
\partial(o,j)=i\}.
\end{equation}
If $V_k(o)=\emptyset$ happens for some $k\geq 1$, then
$V_l(o)=\emptyset$ for all $l\geq k$. With each stratum $V_i$, we
associate a unit vector in $l_2(V)$ defined by
\begin{equation}\label{unitv.}
\ket{\phi_i}=\frac{1}{\sqrt{\kappa_i}}\sum_{k\in V_i(o)}\ket{k}.
\end{equation}
where, $\kappa_i:=|V_i(o)|$ and $\ket{k}$ denotes the eigenket of
$k$-th vertex at the stratum $i$. The closed subspace of $l_2(V)$
spanned by $\{\ket{\phi_i}\}$ is denoted by $\Gamma(G)$. Since
$\{\ket{\phi_i}\}$ becomes a complete orthonormal basis of
$\Gamma(G)$, we often write
\begin{equation}\label{unitv1.}
\Gamma(G)=\sum_k\oplus C\ket{\phi_k}.
\end{equation}
In this stratification for any connected graph $G$ , we have
\begin{equation}\label{dist.}
V_1(\beta)\subseteq V_{i-1}(\alpha)\cup V_i(\alpha)\cup
V_{i+1}(\alpha),
\end{equation}
for each $\beta\in V_i(\alpha)$. Now, recall that the $i$-th
adjacency matrix of a graph $G=(V,E)$ is defined as
\begin{equation}\label{adji.}
\bigl(A_i)_{\alpha, \beta}\;=\;\cases{1 & if $\;
\partial(\alpha,\beta)=i$,\cr 0 & otherwise\cr}\qquad \qquad
(\alpha, \beta \in V).
\end{equation}
Then, for reference state $\ket{\phi_0}$ ($\ket{\phi_0}=\ket{o}$,
with $o\in V$ as reference vertex), we have
\begin{equation}\label{Foc1}
A_i\ket{\phi_0}=\sum_{\beta\in V_{i}(o)}\ket{\beta}.
\end{equation}
Then by using (\ref{unitv.}) and (\ref{Foc1}), we have
\begin{equation}\label{Foc2}
A_i\ket{\phi_0}=\sqrt{\kappa_i}\ket{\phi_i}.
\end{equation}
\subsection{QD type graphs}
Let $A$ be the adjacency matrix of a graph $G=(V,E)$. According to
the stratification (\ref{str.}), we define three matrices $A_+,
A_-$ and $A_0$ as follows: for $i\in V_k$ we set
\begin{equation}\label{A+}
A_+\ket{i}=\sum_{j\in V_{k+1}}\ket{j},\;\;\ A_-\ket{i}=\sum_{j\in
V_{k-1}}\ket{j},\;\;\ A_0\ket{i}=\sum_{j\in V_{k}}\ket{j}.
\end{equation}
for $j \sim i$. Since $i\in V_k$ and $i \sim j$, then $j\in
V_{k-1}\cup V_k \cup V_{k+1}$, where we tacitly understand that
$V_{-1}=\emptyset$. One can easily verify that
$$
(A_+)^*=A_-,\;\;\;\ (A_0)^*=A_0,\;\;\ \mathrm{and}$$
\begin{equation}\label{obata}
A=A_++A_-+A_0
\end{equation}
This is called quantum decomposition of $A$ associated with the
stratification (\ref{str.}). The vector state corresponding to
$\ket{o}=\ket{\phi_0}$, with $o\in V$ as the fixed origin, is
analogous to the vacuum state in Fock space. According to
Ref.\cite{obata}, the $\langle A^m \rangle$ coincides with the
number of $m$-step walks starting and terminating at $o$, also, by
lemma 2.2 of \cite{obata}, if $\Gamma(G)$ is invariant under the
quantum components $A_+,A_-$ and $A_0$, then there exist two
sequences $\{\omega_k\}_{k=1}^{\infty}$ and
$\{\alpha_k\}_{k=1}^{\infty}$ derived from $A$, such that
$$
A_+\ket{\phi_k}=\sqrt{\omega_{k+1}}\ket{\phi_{k+1}},\;\ k\geq0,$$
$$
A_-\ket{\phi_k}=\sqrt{\omega_{k}}\ket{\phi_{k-1}},\;\ k\geq1,
$$
\begin{equation}\label{QDP}
A_0\ket{\phi_k}=\alpha_{k+1}\ket{\phi_{k}},\;\ k\geq0
\end{equation}
(for more details see \cite{js,obata}). Following Ref.\cite{js}, we
will refer to the graphs with this property as QD type graphs and
the parameters $\omega_{k}$ and $\alpha_{k}$ will be called QD
parameters of the graph. We note that in QD type graphs, the
stratification is independent of the choice of reference state.
\subsection{Stieltjes transform of spectral measure associated with
the graph} It is well known that, for any pair $(A,\ket{\phi_0})$ of
a matrix $A$ and a vector $\ket{\phi_0}$, it can be assigned a
measure $\mu$ as follows
\begin{equation}\label{sp1}
\mu(x)=\braket{ \phi_0}{E(x)|\phi_0},
\end{equation}
 where
$E(x)=\sum_i|u_i\rangle\langle u_i|$ is the operator of projection
onto the eigenspace of $A$ corresponding to eigenvalue $x$, i.e.,
\begin{equation}
A=\int x E(x)dx.
\end{equation}
It is easy to see that, for any polynomial $P(A)$ we have
\begin{equation}\label{sp2}
P(A)=\int P(x)E(x)dx,
\end{equation}
where for discrete spectrum the above integrals are replaced by
summation. Therefore, using the relations (\ref{sp1}) and
(\ref{sp2}), the expectation value of powers of adjacency matrix $A$
over starting site $\ket{\phi_0}$ can be written as
\begin{equation}\label{v2}
\braket{\phi_{0}}{A^m|\phi_0}=\int_{R}x^m\mu(dx), \;\;\;\;\
m=0,1,2,....
\end{equation}
The existence of a spectral distribution satisfying (\ref{v2}) is a
consequence of Hamburger's theorem, see e.g., Shohat and Tamarkin
[\cite{st}, Theorem 1.2].

Obviously relation (\ref{v2}) implies an isomorphism from the
Hilbert space of the stratification onto the closed linear span of
the orthogonal polynomials with respect to the measure $\mu$. From
(\ref{dispoly}) and (\ref{Foc2}), we have for distance-regular
graphs
\begin{equation}\label{xx}
\ket{\phi_i}= P'_i(A)\ket{\phi_0},
\end{equation}
where, $P'_i=\frac{1}{\sqrt{\kappa_i}}P_i$ is a polynomial with
real coefficients and degree $i$. Now, substituting (\ref{xx}) in
(\ref{trt}), we get the following  three term recursion relations
for the polynomials $P'_j(x)$
\begin{equation}\label{trt0}
xP'_{k}(x)=\beta_{k+1}P'_{k+1}(x)+\alpha_kP'_{k}(x)+\beta_kP'_{k-1}(x)
\end{equation}
for $k=0,...,d-1$, with $P'_0(x)=1$. Multiplying (\ref{trt0}) by
$\beta_1...\beta_k$, we obtain
\begin{equation}\label{op'}
\beta_1...\beta_kxP'_{k}(x)=\beta_1...\beta_{k+1}P'_{k+1}(x)+\alpha_k\beta_1...\beta_kP'_{k}(x)+\beta_k^2.\beta_1...\beta_{k-1}P'_{k-1}(x).
\end{equation}
By rescaling $P'_k$ as $Q_k=\beta_1...\beta_kP'_k$, the spectral
distribution $\mu$ under question is characterized by the property
of orthonormal polynomials $\{Q_k\}$ defined recurrently by
$$ Q_0(x)=1, \;\;\;\;\;\
Q_1(x)=x,$$
\begin{equation}\label{op}
xQ_k(x)=Q_{k+1}(x)+\alpha_{k}Q_k(x)+\beta_k^2Q_{k-1}(x),\;\;\ k\geq
1.
\end{equation}

If such a spectral distribution is unique, the spectral distribution
$\mu$ is determined by the identity
\begin{equation}\label{sti}
G_{\mu}(x)=\int_{R}\frac{\mu(dy)}{x-y}=\frac{1}{x-\alpha_0-\frac{\beta_1^2}{x-\alpha_1-\frac{\beta_2^2}
{x-\alpha_2-\frac{\beta_3^2}{x-\alpha_3-\cdots}}}}=\frac{Q_{d-1}^{(1)}(x)}{Q_{d}(x)}=\sum_{l=0}^{d-1}
\frac{A_l}{x-x_l},
\end{equation}
where, $x_l$ are the roots  of polynomial $Q_{d}(x)$. $G_{\mu}(z)$
is called the Stieltjes function (Stieltjes/Hilbert transform of
spectral distribution $\mu$) and polynomials $\{Q_{k}^{(1)}\}$ are
defined recurrently as
$$Q_{0}^{(1)}(x)=1, \;\;\;\;\;\
    Q_{1}^{(1)}(x)=x-\alpha_1,$$
\begin{equation}\label{oq}
xQ_{k}^{(1)}(x)=Q_{k+1}^{(1)}(x)+\alpha_{k+1}Q_{k}^{(1)}(x)+\beta_{k+1}^2Q_{k-1}^{(1)}(x),\;\;\
k\geq 1,
\end{equation}
respectively. Then, from (\ref{sti}), one can see that the
spectral distribution $\mu$ is written as
\begin{equation}\label{spdis}
\mu(x)=\sum_{l=0}^{d-1} A_l\delta(x-x_l).
\end{equation}
The coefficients $A_l$ appearing in (\ref{sti}) and (\ref{spdis})
are calculated as
\begin{equation}\label{Gauss}
A_l=\lim_{x\rightarrow x_l}(x-x_l)G_{\mu}(x).
\end{equation}
(for more details see Refs.\cite{st, obh,tsc,obah}.)
\section{Continuous-time quantum walk on an arbitrary graph} For a Given
undirected graph $\Gamma$ with $n$ vertices and adjacency matrix
$A$, one can define the Laplacian of $\Gamma$ as $L=A-D$, where $D$
is the diagonal matrix with $D_{jj}=deg(j)$, the degree of vertex
$j$. Classically, suppose that $P(t)$ is a time-dependent
probability distribution of a stochastic (particle) process on
$\Gamma$. The classical evolution of the continuous-time walk is
given by the Kolmogorov equation
\begin{equation}\label{master}
\frac{dP(t)}{dt}=LP(t).
\end{equation}
The solution to this equation, modulo some conditions, is
$P(t)=e^{tL}P(0)$, which can be solved by diagonalizing the
symmetric matrix $L$. This spectral approach requires full knowledge
of the spectrum of $L$.

A quantum analogue of the classical walk, the so-called CTQW, uses
the Schr\"{o}dinger equation in place of the Kolmogorov equation,
where $L$ is chosen as the Hamiltonian of the walk. This is because
we can view $L$ as the generator matrix that describes an
exponential distribution of waiting times at each vertex. Let
$\ket{\phi(t)}$ be a time-dependent amplitude of the quantum process
on $\Gamma$. Then, the wave evolution of the quantum walk is
governed by
\begin{equation}\label{master3}
i\hbar\frac{d}{dt}\ket{\phi(t)}=H\ket{\phi(t)}.
\end{equation}
Assuming $\hbar = 1$ for simplicity, the solution to (\ref{master3})
is given by $\ket{\phi(t)} = e^{-iHt} \ket{\phi(0)}$ which, again,
is solvable via spectral techniques.  On $d$-regular graphs, we have
$D = \frac{1}{d}I$, and since $A$ and $D$ commute, we get
\begin{equation} \label{eqn:phase-factor}
e^{-itH} = e^{-it(A-\frac{1}{d}I)} = e^{it/d}e^{-itA}.
\end{equation}
This introduces an irrelevant phase factor in the wave evolution.
Hence we can consider $H=A=A_1$. Thus, we have
\begin{equation}
\ket{\phi(t)}=e^{-iAt}\ket{\phi(0)}.
\end{equation}

In the case of distance regular graphs, according to
(\ref{dispoly}) the adjacency matrices are polynomial functions of
$A$, hence by using (\ref{v2}) and (\ref{xx}), the matrix elements
$\label{cw1} \braket{\phi_{l}}{A^m\mid \phi_0}$ for $m=0,1,...$
can be written as
\begin{equation}\label{cw1}
\braket{\phi_{l}}{A^m\mid \phi_0}=\braket{\phi_{0}}{P'_{l}(A)
A^m\mid \phi_0}=\int_{R}x^{m}P'_{l}(x)\mu(dx).
\end{equation}

By using (\ref{cw1}), the probability amplitude of observing the
walk at $l$-th stratum at time $t$ can be evaluated as
\begin{equation} \label{v4}
q_l(t)\equiv\braket{\phi_{l}}{e^{-iAt}|\phi_0}=\int_{R}e^{-ixt}P'_{l}(x)\mu(dx)=\frac{1}{\sqrt{\omega_1...\omega_l}}\int_{R}e^{-ixt}Q_{l}(x)\mu(dx).
\end{equation}
In particular, the probability of observing the walk at starting
site at time $t$ is given by
\begin{equation} \label{Q0}
q_0(t)=\braket{\phi_{0}}{e^{-iAt}|\phi_0}=\int_{R}e^{-ixt}\mu(dx).
\end{equation}
The conservation of probability $\sum_{l=0}{\mid
\braket{\phi_{l}}{\phi_0(t)}\mid}^2=1$ follows immediately from
(\ref{v4}) by using the completeness relation of orthogonal
polynomials $P_l(x)$.

We notice that the formula (\ref{v4}) indicates a canonical
isomorphism between the interacting Fock space of CTQW on distance
regular graphs and the closed linear span of the orthogonal
polynomials generated by recursion relations (\ref{dra}). This
isomorphism was meant to be, a reformulation of CTQW (on distance
regular graphs), which describes quantum states by polynomials
(describing quantum state $\ket{\phi_k}$ by $P_k(x)$), and  make a
correspondence between functions of operators ($q$-numbers) and
functions of classical quantity ($c$-numbers), such as the
correspondence between $e^{-iAt}$ and $e^{-ixt}$. This isomorphism
is similar to the isomorphism between Fock space of annihilation
and creation operators $a$, $a^{\dag}$ with space of functions of
coherent states' parameters in quantum optics, or the isomorphism
between Hilbert space of momentum and position operators, and
spaces of function defined on phase space in Wigner function
formalism.

According to the result of Ref. \cite{js}, the walk has the same
probability amplitudes at all sites belonging to the same stratum
and so the evaluation of $q_l(t)$ leads to the determination of the
amplitudes at sites belonging to the $l$-th stratum $V_l(o)$. Then,
by using (\ref{unitv.}) we have
\begin{equation} \label{v44}
q_l(t)=\braket{\phi_{l}}{e^{-iAt}|\phi_0}=\frac{1}{\sqrt{\kappa_l}}\sum_{\alpha\in
V_l(o)}\langle\alpha|e^{-iAt}|o\rangle=\sqrt{\kappa_l}\langle\alpha|e^{-iAt}|o\rangle.
\end{equation}
Therefore, the probability amplitude of observing the walk at the
site $\alpha\in V_l(o)$ at time $t$ is given by
\begin{equation} \label{v444}
p_{\alpha}(t)\equiv\langle\alpha|e^{-iAt}|o\rangle=\frac{1}{\sqrt{\kappa_l}}q_l(t)\;\
,\;\ \mbox{for}\;\ \mbox{all}\;\;\ \alpha\in V_l(o).
\end{equation}
\subsection{Evaluation of probability amplitudes by
using Laplace transform} In this section, we give a procedure for
calculating the probability amplitudes of CTQWs on graphs by using
the Laplace transformation. To do so, first we give the definition
of the Laplace transform. The Laplace transform of a
time-dependent function $f(t)$, denoted by
$\hat{f}(s)=\textit{L}{f(t)}$, is defined as
\begin{equation} \label{laplacet}
\hat{f}(s)=\int_{0}^{\infty}e^{-st}f(t)dt.
\end{equation}
The Laplace transform  has the following basic properties :\\
$\mbox{Linearity}:
\textit{L}\{af(t)+bg(t)\}=a\hat{f}(s)+b\hat{g}(s),$\\
$\mbox{Derivative}: \textit{L}{f'(t)}= s \hat{f}(s)-f(0),$\\
$\mbox{Shifting}:\textit{ L}{e^{at}f(t)}=\hat{f}(s-a)$.\\ (for
more information related to the Laplace transform see
\cite{3},\cite{lap1} and \cite{lap2}).

By taking the Laplace transform of the amplitudes $q_l(t)$ in
(\ref{v4}) and using the equation (\ref{spdis}), one can obtain
$$\hat{q}_l(s)=\frac{1}{\sqrt{\omega_1...\omega_l}}\int_{0}^{\infty}e^{-st}\int_{R}e^{-ixt}Q_l(x)\mu(dx)dt=\frac{1}{\sqrt{\omega_1...\omega_l}}\sum_iA_iQ_l(x_i)\int_{0}^{\infty}e^{-(s+ix_i)t}dt=$$
\begin{equation} \label{ampl}
\frac{1}{\sqrt{\omega_1...\omega_l}}\sum_i\frac{A_iQ_l(x_i)}{s+ix_i}=
\frac{i}{\sqrt{\omega_1...\omega_l}}\sum_i\frac{A_iQ_l(x_i)}{is-x_i}.
\end{equation}
Therefore, although it is not an easy work in the most cases, the
probability amplitudes $q_l(t)$ of the walk can be obtained by
taking the inverse Laplace transform of (\ref{ampl}), but we have
no need to do so. Instead, we calculate only $\hat{q}_0(s)$ as
follows
\begin{equation} \label{amp0}
\hat{q}_0(s)=\int_{0}^{\infty}e^{-st}\int_{R}e^{-ixt}\mu(dx)dt=i\sum_i\frac{A_i}{is-x_i}=iG_{\mu}(is).
\end{equation}
and obtain the probability  amplitude of observing the walk at
starting site at time $t$ as
\begin{equation} \label{amp00}
q_0(t)=i\textit{L}^{-1}(G_{\mu}(is)).
\end{equation}

From the formulas (\ref{v4}) and (\ref{Q0}), it can be easily seen
that
\begin{equation} \label{Qll}
q_l(t)=P'_l(i\frac{d}{dt})q_0(t)=\frac{1}{\sqrt{\omega_1...\omega_l}}Q_l(i\frac{d}{dt})q_0(t).
\end{equation}
In fact, we need only to know the probability amplitude of
observing the walk at starting site at time $t$ and the
polynomials $Q_l(x)$ which are obtained via the recursion
relations (\ref{op}). Then, by using the equations (\ref{amp00})
and (\ref{Qll}), the probability amplitudes $q_l(t)$ can be
evaluated.

It could be noticed that, in this approach we do not any need to
take the inverse Stieltjes transform to obtain the spectral
measure $\mu$; Instead we take the inverse Laplace transform of
the Stieltjes function $G_{\mu}(is)$ which is more popular and
convenient (there are many handbooks and tables of Laplase
transforms \cite{3},\cite{lap1},\cite{lap2} and computer
programmings for this purpose) in comparison with the inverse
Stieltjes transform.
 In the following section, we calculate the
probability amplitudes for CTQW on some important QD graphs, i.e.,
distance-regular graphs and some non-QD type graphs, explicitly.
\section{Examples}
In this section we consider CTQW on some examples of graphs and
calculate the corresponding probability amplitudes of the walk on
them by employing the techniques introduced in the previous
section.
\subsection{Examples of distance-regular graphs}
In this subsection first we consider some set of important QD
graphs called distance-regular graphs. To do so, we recall the
definitions and properties related to distance-regular graphs:\\
\textbf{Definition.} An undirected connected graph $G=(V,E)$ is
called distance-regular graph (DRG) with diameter $d$ if it
satisfies the following distance-regularity condition:\\
For all $h,i,j\in\{0,1,...,d\}$, and $\alpha,\beta$ with
$\partial(\alpha,\beta)=h$, the number
\begin{equation}
p^k_{ij}=\mid \{\gamma\in V : \partial(\alpha, \beta)=i \;\ and
\;\ \partial(\gamma,\beta)=j\}\mid
\end{equation}
is constant in that it depends only on $h, i, j$ but does not
depend on the choice of $\alpha$ and $\beta$. Then,
$p^h_{ij}:=|V_i(\alpha)\cap V_j(\beta)|$ for all $\alpha,\beta\in
V$ with $\partial(\alpha,\beta)=h$. This number is called the
intersection number. In a distance regular graph, $p_{j1}^i=0 $
(for $i\neq 0$, $j$ is not $\{i-1, i, i+1 \}$). The intersection
numbers  of the graph are defined as
\begin{equation}\label{abc}
 a_i=p_{i1}^i, \;\;\;\  b_i=p_{i+1,1}^i, \;\;\;\
 c_i=p_{i-1,1}^i\;\ .
\end{equation}
These intersection numbers and the valencies $\kappa_i$ satisfy
the following obvious conditions
$$a_i+b_i+c_i=\kappa,\;\;\ \kappa_{i-1}b_{i-1}=\kappa_ic_i ,\;\;\
i=1,...,d,$$
\begin{equation}\label{intersec}
\kappa_0=c_1=1,\;\;\;\ b_0=\kappa_1=\kappa, \;\;\;\ (c_0=b_d=0).
\end{equation}
Thus, all parameters of the graph can be obtained from the
intersection array $\{b_0,...,b_{d-1};c_1,...,c_d\}$.

It could be noticed that, for adjacency matrices of a distance
regular graph, we have
$$
A_1A_i=b_{i-1}A_{i-1}+a_iA_i+c_{i+1}A_{i+1}, \;\;\;\ \mbox{for}
\;\ i=1,2,...,d-1,
$$
\begin{equation}\label{dra}
A_1A_d=b_{d-1}A_{d-1}+a_dA_d.
\end{equation}
Using the recursion relations (\ref{dra}), one can show that $A_i$
is a polynomial in $A_1$ of degree $i$, i.e., we have
\begin{equation}\label{dispoly}
A_i=P_i(A_1), \;\;\ i=1,2,...,d ,
\end{equation}
and conversely $A_1^i$ can be written as a linear combination of
$I, A_1, ..., A_d$ (for more details see for example \cite{js}).

It should be noticed that, for distance-regular graphs, the unit
vectors $\ket{\phi_i}$ for $i=0,1,...,d$ defined as in
(\ref{unitv.}), satisfy the following three-term recursion
relations
\begin{equation}\label{trt}
A\ket{\phi_i}=\beta_{i+1}\ket{\phi_{i+1}}+\alpha_i\ket{\phi_i}+\beta_{i}\ket{\phi_{i-1}},
\end{equation}
where, the coefficients $\alpha_i$ and $\beta_i$ are defined as
\begin{equation}\label{omegal}
\alpha_k=\kappa-b_{k}-c_{k},\;\;\;\;\
\omega_k\equiv\beta^2_k=b_{k-1}c_{k},\;\;\ k=1,...,d,
\end{equation}
i.e., in the basis of unit vectors $\{\ket{\phi_i},i=0,1,...,d\}$,
the adjacency matrix $A$ is projected to the following symmetric
tridiagonal form:
\begin{equation}\label{trid}
A=\left(
\begin{array}{cccccc}
 \alpha_0 & \beta_1 & 0 & ... &...&0 \\
      \beta_1 & \alpha_1 & \beta_2 & 0 &...&0 \\
      0 & \beta_2 & \alpha_3 & \beta_3 & \ddots&\vdots \\
     \vdots & \ddots &\ddots& \ddots &\ddots &0\\
     0 & \ldots  &0 &\beta_{d-1} & \alpha_{d-1} &\beta_{d}\\
          0&... & 0 &0 & \beta_{d} & \alpha_{d}\\
\end{array}
\right).
\end{equation}

In the following, we consider CTQW on some examples of
distance-regular graphs in details, where in the Appendix $A$, we
give a table of important distance-regular graphs together with
the corresponding probability amplitude $q_0(t)$.
\subsubsection{Complete graph $K_n$} The complete graph $K_n$ is
the simplest example of distance-regular graphs. This graph has
$n$ vertices with $n(n-1)/2$  edges, the degree of each vertex is
$\kappa=n-1$ also the graph has diameter $d=1$. The intersection
array of the graph is $\{b_0;c_1\}=\{n-1;1\}$. Then, the graph has
only two strata $V_0(\alpha)=\alpha$ and
$V_1(\alpha)=\{\beta:\beta\neq\alpha\}$ with QD parameters
$\{\alpha_1;\omega_1\}=\{n-2,n-1\}$. By using (\ref{sti}), the
Stieltjes function is calculated as
\begin{equation}\label{sticom}
G_{\mu}(x)=\frac{x-n+2}{x^2-(n-2)x-n+1}.
\end{equation}
From (\ref{amp0}) and (\ref{sticom}), we obtain
\begin{equation}\label{sticom1}
\hat{q}_0(s)=\frac{s+i(n-2)}{s^2+is(n-2)+n-1}=\frac{1}{n}\{\frac{1}{s+i(n-1)}+\frac{n-1}{s-i}\}.
\end{equation}
Therefore, by taking the inverse Laplace transform of
(\ref{sticom1}), we obtain the probability amplitude of observing
the walk at starting site at time $t$ as follows
\begin{equation}\label{sticom2}
q_0(t)=\textit{L}^{-1}(\hat{q}_0(s))=\frac{1}{n}(e^{-i(n-1)t}+(n-1)e^{it}).
\end{equation}
For calculating the probability amplitude $q_1(t)$, first recall
that $\kappa_1=n-1$ and $P'_1(x)=\frac{x}{\sqrt{n-1}}$, then by
using (\ref{Qll}) we obtain
\begin{equation}\label{sticom3}
q_1(t)=\frac{i}{\sqrt{n-1}}\frac{d}{dt}q_0(t)=\frac{\sqrt{n-1}}{n}(e^{-i(n-1)t}-e^{it}).
\end{equation}
\subsubsection{Strongly regular graphs} For strongly regular graphs
we have three strata and two kinds of two-point resistances
$R_{\alpha\beta^{(1)}}$ and $R_{\alpha\beta^{(2)}}$. The QD
parameters of the graph with parameters $(v,\kappa,\lambda,\mu)$
are given by
\begin{equation}
\alpha_1=\lambda,\;\ \alpha_2=\kappa-\mu;\;\;\ \omega_1=\kappa,\;\
\omega_2=\mu(\kappa-\lambda-1).
\end{equation}
Then, by using (\ref{sti}) and (\ref{amp0}), one can obtain
\begin{equation}\label{stielts}
\hat{q}_0(s)=\frac{s^2+i(\kappa+\lambda-\mu)s-\kappa(\lambda-\mu)-\mu}{s^3+i(\kappa+\lambda-\mu)s^2+(\kappa(\mu-\lambda+1)-\mu)s+i\kappa(\kappa-\mu)}.
\end{equation}
Therefore, the probability amplitude $q_0(t)$ can be obtained as
follows
\begin{equation}\label{stieltsq00}
q_0(t)=A_1e^{-i\kappa
t}+A_2e^{-\frac{i}{2}(\lambda-\mu+\sqrt{(\lambda-\mu)^2-4(\mu-\kappa)})t}+A_3e^{-\frac{i}{2}(\lambda-\mu+\sqrt{(\lambda-\mu)^2-4(\mu-\kappa)})t},
\end{equation}
where, $$ A_1=\frac{\mu}{\kappa^2-\kappa(\lambda-\mu)+(\mu-\kappa)},
$$
$$
A_2=\frac{-\kappa\sqrt{(\lambda-\mu)^2-4(\mu-\kappa)}+\kappa(\lambda-\mu)+2\kappa}
{(\lambda-\mu-2\kappa)\sqrt{(\lambda-\mu)^2-4(\mu-\kappa)}+(\lambda-\mu)^2-4(\mu-\kappa)},
$$
\begin{equation}\label{stron3}
A_3=\frac{\kappa\sqrt{(\lambda-\mu)^2-4(\mu-\kappa)}+\kappa(\lambda-\mu)+2\kappa}
{(-\lambda+\mu+2\kappa)\sqrt{(\lambda-\mu)^2-4(\mu-\kappa)}+(\lambda-\mu)^2-4(\mu-\kappa)}.
\end{equation}

In the remaining part of this example, we study CTQW on the
following two well-known strongly regular graphs:\\
\textbf{A. Petersen graph}\\ The Petersen  graph \cite{Ass.sch.}
is a strongly regular graph with parameters $(v, \kappa, \lambda,
\mu)=(10,3,0,1)$, the intersection array
$\{b_0,b_{1};c_1,c_2\}=\{3,2;1,1\}$ and QD parameters
$\{\alpha_1,\alpha_2;\omega_1,\omega_2\}=\{0,2;3,2\}$. Then, by
using (\ref{stron3}), we obtain
\begin{equation}\label{stron33'}
A_1=\frac{1}{10},\;\;\ A_2=\frac{1}{2},\;\;\ A_3=\frac{2}{5}.
\end{equation}
By substituting (\ref{stron33'}) in (\ref{stieltsq00}), we obtain
the probability amplitude of observing the walk at starting site
at time $t$ as follows
\begin{equation}\label{stieltp00}
q_0(t)=\frac{1}{10}(5e^{-it}+4e^{2it}+e^{-3it}).
\end{equation}
Then the probability amplitudes $q_1(t)$ and $q_2(t)$ are easily
calculated as
$$q_1(t)=\frac{i}{\sqrt{3}}\frac{d}{dt}q_0(t)=\frac{1}{\sqrt{3}}(\frac{1}{2}e^{-it}-\frac{4}{5}e^{2it}+\frac{3}{10}e^{-3it}),$$
\begin{equation}\label{stieltp000}
q_2(t)=\frac{1}{\sqrt{6}}(-\frac{d^2}{dt^2}-3)q_0(t)=\frac{1}{\sqrt{6}}(-e^{-it}+\frac{2}{5}e^{2it}+\frac{2}{5}e^{-3it}).
\end{equation}
\textbf{B. Normal subgroup scheme}\\ \textbf{Definition } The
partition $P=\{P_0,P_1,...,P_d \}$ of
a finite group $G$ is called a blueprint if\\
(i) $P_0=\{e\}$\\(ii) for i=1,2,...,d if $g\in P_i$ then
$g^{-1}\in P_i$\\(iii) the set of relations
$R_i=\{(\alpha,\beta)\in G\otimes G|\alpha^{-1}\beta\in P_i\}$ on
$G$ form an association scheme\cite{Ass.sch.}. The set of real
conjugacy classes given in Appendix $A$ of Ref. \cite{js} is an
example of blueprint on $G$. Also, one can show that in the
regular representation, the class sums $\bar{P_i}$ for
$i=0,1,...,d$ defined  as
\begin{equation}
\bar{P_i} = \sum_{\gamma\in P_i}\gamma \in CG, \;\;\ i=0,1,...,d,
\end{equation}
are the adjacency matrices of a blueprint scheme.

In Ref.\cite{js}, it has been shown that, if $H$ be a normal
subgroup of $G$, the following blueprint classes
\begin{equation}\label{xxx}
P_0=\{e\}, \;\;\;\ P_1=G-\{H\}, \;\;\;\ P_2=H -\{e\},
\end{equation}
define a strongly regular graph with parameters $(v, \kappa,
\lambda, \mu)=(g,g-h,g-2h,g-h)$ and the following intersection
array
\begin{equation}
\{b_0,b_{1};c_1,c_2\}=\{g-h,h-1;1,g-h\},
\end{equation}
 where, $g:=|G|$ and $h:=|H|$. It is interesting to note that in normal subgroup scheme, the
intersections numbers and other parameters depend only on the
cardinalities of the group and its normal subgroup. By using
(\ref{omegal}), the QD parameters are given by
$\{\alpha_1,\alpha_2;\omega_1,\omega_2\}=\{g-2h,0;g-h,(g-h)(h-1)\}$.

As an example, we consider the dihedral group $G=D_{2m}$, where
its normal subgroup is $H=Z_m$. Therefore, the blueprint classes
are given by
\begin{equation}
P_0=\{e\}, \;\;\ P_1=\{b,ab,a^2b,...,a^{m-1}b\}, \;\;\
P_2=\{a,a^{2},...,a^{(m-1)}\},
\end{equation}
which form  a strongly regular graph with parameters $(2m,m,0,m)$
 and the following intersection numbers and  QD parameters
\begin{equation}
\{b_0,b_{1};c_1,c_2\}=\{m,m-1;1,m\};\;\
\{\alpha_1,\alpha_2;\omega_1,\omega_2\}=\{0,0;m,m(m-1)\}.
\end{equation}

Then, by using (\ref{stron3}), we obtain
\begin{equation}\label{stron33}
A_1=\frac{1}{2m},\;\;\ A_2=\frac{m-1}{m},\;\;\ A_3=\frac{1}{2m}.
\end{equation}
Now, by substituting (\ref{stron33}) in (\ref{stieltsq00}), we
obtain the probability amplitude of observing the walk at starting
site at time $t$ as follows
\begin{equation}\label{stieltp00xx}
q_0(t)=\frac{1}{m}(m-1+\cos mt).
\end{equation}
The other probability amplitudes are calculated as
$$q_1(t)=\frac{i}{\sqrt{m}}\frac{d}{dt}q_0(t)=-\frac{i}{\sqrt{m}}\sin
mt,$$
\begin{equation}\label{stieltp00xxx}
q_2(t)=\frac{1}{\sqrt{m}}(-\frac{d^2}{dt^2}-m\sqrt{m})q_0(t)=(\sqrt{m}-1)\cos
mt-(m-1).
\end{equation}
\subsubsection{Cycle graph $C_{2m}$}
The cycle graph or cycle with $n$ vertices is denoted by $C_n$
with $\kappa=2$. We consider $n=2m$ (the case $n=2m+1$ can be
considered similarly). The intersection array is given by
\begin{equation}\label{intcye.}
\{b_0,...,b_{m-1};c_1,...,c_m\}=\{2,1,...,1,1;1,...,1,2\}
\end{equation}
Then, by using (\ref{omegal}), the QD parameters are given by
\begin{equation}\label{QDcye.}
\alpha_i=0, \;\ i=0,1,...,m; \;\ \omega_1=\omega_m=2,\;\
\omega_i=1,\;\ i=2,...,m-1,
\end{equation}
Therefore, by using (\ref{sti}) and (\ref{amp0}), one can obtain
\begin{equation}
\hat{q}_0(s)=\frac{i}{n}
\frac{T'_n(\frac{is}{2})}{T_n(\frac{is}{2})}.
\end{equation}
Then, the probability amplitude $q_0(t)$ is obtained as
\begin{equation}
q_0(t)=\frac{1}{n}(\cos 2t+\sum_{l=1,l\neq n}^{2n-1}e^{-2it\cos
\frac{2l\pi}{2n}}).
\end{equation}
From (\ref{Qll}), one can calculate
\begin{equation}\label{cycle1}
q_1(t)=\frac{1}{\sqrt{2}}\frac{id}{dt}(q_0(t))=-\frac{\sqrt{2}i}{n}(\sin
2t+i\sum_{l=1,l\neq n}^{2n-1}\cos
\frac{2l\pi}{2n}e^{-2it\cos\frac{2l\pi}{2n}}).
\end{equation}
The amplitudes $q_l(t)$, for $l>1$ can be calculated similarly,
where the results thus obtained are in agreement with those of
Ref.\cite{Ahmadi}.

It could be noticed that, in the limit of the large $n$, the cycle
graph tend to the infinite line graph $Z$ and the Stieltjes
function reads as
\begin{equation}\label{st.2}
G_{\mu}(x)=\frac{1}{\sqrt{x^2-4}}.
\end{equation}
Therefore, by using (\ref{amp0}) we obtain $q_0(t)$ as follows
\begin{equation}
q_0(t)=L^{-1}(iG_{\mu}(is))=J_0(2t),
\end{equation}
where, the $J_0$ is Bessel function. From (\ref{Qll}), we can
calculate
$$q_1(t)=\frac{1}{\sqrt{2}}\frac{id}{dt}(J_0(2t))=-\sqrt{2}iJ_1(2t),$$
\begin{equation}\label{stiinf2}
q_2(t)=\frac{1}{\sqrt{2}}(-\frac{d^2}{dt^2}-2)(J_0(2t))=-\sqrt{2}J_2(2t).
\end{equation}
By using (\ref{stiinf2}), one can deduce that
\begin{equation}\label{stiinff}
q_l(t)=\sqrt{2}(-i)^lJ_l(2t),
\end{equation}
where the results are in agreement with those of Ref.\cite{konno}.
\subsubsection{Johnson graphs} Let $n\geq 2$ and $d\leq n/2$. The
Johnson graph $J(n,d)$ has all $d$-element subsets of
$\{1,2,...,n\}$ such that two $d$-element subsets are adjacent if
their intersection has size $d-1$. Two $d$-element subsets are
then at distance $i$ if and only if they have exactly $d-i$
elements in common. The Johnson graph $J(n,d)$ has
$v=\frac{n!}{d!(n-d)!}$ vertices, diameter $d$ and the valency
$\kappa=d(n-d)$. Its intersection array is given by
\begin{equation}
b_i=(d-i)(n-d-i), \;\;\;\ 0\leq i\leq d-1; \;\;\ c_i=i^2, \;\;\;\
1\leq i\leq d.
\end{equation}
Then, by using (\ref{omegal}) the QD parameters are given by
\begin{equation}
\alpha_k=k(n-2k),\;\;\ \omega_k=k^2(d-k+1)(n-d-k+1).
\end{equation}
For $d=2$, the Stieltjes function is calculated as
\begin{equation}\label{sd2}
G_{\mu}(x)=\frac{x-n+2}{x^2-(n-2)x-2(n-2)}.
\end{equation}
Then by using (\ref{amp0}) and (\ref{sticom}), we obtain
\begin{equation}\label{sticom1}
\hat{q}_0(s)=\frac{s+i(n-2)}{s^2+is(n-2)-2(n-2)}=\frac{1}{2}\{\frac{1-\sqrt{\frac{n-2}{n+6}}}{s+i(\frac{n-2+\sqrt{(n-2)(n+6)}}{2})}+\frac{1+\sqrt{\frac{n-2}{n+6}}}{s+i(\frac{n-2-\sqrt{(n-2)(n+6)}}{2})}\}.
\end{equation}
Then, the probability amplitude $q_0(t)$ can be obtained  as
follows
\begin{equation}\label{stiTchq01}
q_0(t)=L^{-1}(iG_{\mu}(is))=e^{-i\frac{n-2}{2}t}\{\cos(\frac{\sqrt{(n-2)(n+6)}}{2})t+i\sqrt{\frac{n-2}{n+6}}\sin(\frac{\sqrt{(n-2)(n+6)}}{2})t\}.
\end{equation}
\subsection{Examples of non-QD type graphs}
\subsubsection{Tchebichef graphs} By choosing Tchebichef
polynomials of the first kind (the second kind )with scaling
factor $1/2^m$ as orthogonal polynomials appearing in the
recursion relations (\ref{op}), i.e.,
$Q_n(x)=2^{(m-1)n+1}T_n(x/2^m)$ $(2^{(m-1)n}U_n(x/2^m))$, one can
obtain a class of finite and infinite graphs of Tchebichef type,
with QD parameters $\omega_1 = 2^{2(m-1)+1}$, $\omega_k =
2^{2(m-1)}$, $k = 2, 3, ...$, and $\alpha_k = 0$, for $k = 1, 2,
3, ...$ $(\alpha_k = 0;\omega_k = 2^{2(m-1)}$, $k = 1, 2, ...)$.
Then, we obtain
\begin{equation}\label{stiTch}
\hat{q}_0(s)=\frac{i}{n}
\frac{T'_n(\frac{is}{2^m})}{T_n(\frac{is}{2^m})}\;\
(\frac{i}{2^{m-1}}\frac{U_n(\frac{is}{2^m})}{U_{n+1}(\frac{is}{2^m})}).
\end{equation}
Therefore, the probability amplitude $q_0(t)$ can be obtained as
follows
\begin{equation}\label{stiTchq01}
q_0(t)=\frac{1}{n}\sum_{l=0}^{n-1}e^{-i2^mt\cos
\frac{(2l+1)\pi}{2n}}
\end{equation}
for the first kind and
\begin{equation}\label{stiTchq02}
q_0(t)=\frac{2}{n+2}\sum_{l=0}^{n-1}\sin^2
\frac{l\pi}{n+2}e^{-i2^mt\cos \frac{l\pi}{n+2}}
\end{equation}
for the second kind.
\subsubsection{Finite path graph $P_n$}
For $m=1$ in (\ref{stiTch}) and Tchebishef polynomials of the
second kind we obtain finite path graph $P_n= \{0,1,2,...,n-1\}$,
where it is a $n$-vertex graph with $n-1$ edges all on a single
open path \cite{js}. For this graph, the stratification depends on
the choice of the starting site of walk. If we choose the first
vertex as the starting site of the walk, the QD parameters are
given by $\omega_i=1,\;\;\ \alpha_i=0,\;\;\ i=1,...,n-1$. Then by
using (\ref{stiTchq02}) we have
\begin{equation}\label{stitch111}
q_0(t)=\frac{2}{n+2}\sum_{k=1}^{n+1}\sin^2(\frac{k\pi}{n+2})e^{-it\cos(\frac{k\pi}{n+2})}.
\end{equation}
From (\ref{Qll}) and  the recursion relations
$P'_{l+1}(x)=xP'_l(x)-P'_{l-1}(x)$, one can obtain the probability
amplitude of observing the walk at $l$-th stratum at time $t$ as
follows
\begin{equation}\label{stitch1111}
q_l(t)=\frac{2}{n+2}\sum_{k=1}^{n+1}\sin(\frac{k\pi}{n+2})\sin(\frac{(l+1)k\pi}{n+2})e^{-it\cos(\frac{k\pi}{n+2})},\;\;\
l\geq 1.
\end{equation}
 where the results thus obtained are in agreement with those
of Ref.[26].

In the limit of the large $n$, one can calculate the Stieltjes
function as follows
\begin{equation}\label{stiinf}
G_{\mu}(x)=\frac{1}{x-\frac{1}{x-\frac{1}
{x-\frac{1}{x-\cdots}}}}=\frac{1}{x-G_{\mu}(x)},
\end{equation}
then we obtain the following closed form for the Stieltjes function
\begin{equation}\label{stiinf1}
G_{\mu}(x)=\frac{x-\sqrt{x^2-4}}{2}.
\end{equation}
Therefore, by using (\ref{amp0}) we have
\begin{equation}\label{stiinf2}
\hat{q}_0(s)=\frac{\sqrt{s^2+4}-s}{2}
\end{equation}
and then the probability amplitude of  observing the walk at
starting site at time $t$ is given by
\begin{equation}\label{stiinf3}
q_0(t)=J_0(2t)+J_2(2t).
\end{equation}
Now, by using (\ref{Qll}), the probability amplitudes $q_l(t)$ are
obtained as follows
\begin{equation}\label{stitchinf}
q_l(t)=i^l(J_l(2t)+J_{l+2}(2t)).
\end{equation}
\subsubsection{The graphs  $G_n$}
For $m=\frac{3}{2}$ in (\ref{stiTch}) and Tchebishef polynomials
of the second kind we obtain a sequence of graphs denoted by $G_n$
(for more details see Ref.\cite{7}). The number of vertices in
$G_n$ is $2^{n+1}+2^n-2$. In general, $G_n$ consists of two
balanced binary trees of depth $n$ with the $2^n$, $n$-th level
vertices of the two trees pairwise identified. For the quantum
walk on $G_n$, we assume that the starting site of the walk is the
root of a tree and calculate the probability amplitude of the
presence of the walk at the other vertices as a function of time.
Clearly, the graph $G_n$ has $(2n+1)$ strata, where the $j$-th
stratum consists of $2^j$ vertices for $j=1, 2, ...,n+1$ and
$2^{2n+1-j}$ for $j=n+1,..., 2n+1$. The QD parameters are
$\omega_i=2,\;\ \alpha_i=0,\;\ i=1,...,2n$.

By using (\ref{stiTchq02}) we obtain
\begin{equation}\label{stitch111}
q_0(t)=\frac{2}{n+2}\sum_{k=1}^{n+1}\sin^2(\frac{k\pi}{n+2})e^{-i2\sqrt{2}t\cos(\frac{k\pi}{n+2})}.
\end{equation}
From (\ref{Qll}) and  the recursion relations
$P'_{l+1}(x)=xP'_l(x)-2P'_{l-1}(x)$, one can obtain the
probability amplitude of observing the walk at $l$-th stratum at
time $t$ as follows
\begin{equation}\label{stitch1111}
q_l(t)=\frac{2}{n+2}\sum_{k=1}^{n+1}\sin(\frac{k\pi}{n+2})\sin(\frac{(l+1)k\pi}{n+2})e^{-i2\sqrt{2}t\cos(\frac{k\pi}{n+2})},\;\;\
l\geq 1.
\end{equation}

In the limit of the large $n$, one can calculate the Stieltjes
function as follows
\begin{equation}\label{stiinfG}
G_{\mu}(x)=\frac{1}{x-\frac{2}{x-\frac{2}
{x-\frac{2}{x-\cdots}}}}=\frac{1}{x-2G_{\mu}(x)},
\end{equation}
then we obtain the following closed form for the Stieltjes function
\begin{equation}\label{stiinfG1}
G_{\mu}(x)=\frac{x-\sqrt{x^2-8}}{4}.
\end{equation}
Therefore, by using (\ref{amp0}) we have
\begin{equation}\label{stiinfG2}
\hat{q}_0(s)=\frac{\sqrt{s^2+8}-s}{4}
\end{equation}
and then the probability amplitude of  observing the walk at
starting site at time $t$ is given by
\begin{equation}\label{stiinfG3}
q_0(t)=J_0(2\sqrt{2}t)+J_2(2\sqrt{2}t).
\end{equation}
Now, by using (\ref{Qll}), the probability amplitudes $q_l(t)$ are
obtained as follows
\begin{equation}\label{stitchinfG}
q_l(t)=i^l(J_l(2\sqrt{2}t)+J_{l+2}(2\sqrt{2}t)).
\end{equation}
\subsubsection{Finite path graph with the second vertex as starting site of the walk}
Now, we study an example of non-QD type graphs such that as the
authors have shown in Ref. \cite{js2}, one can give the three term
recursion to the graph by using the Krylov-subspace Lanczos
algorithm applied to the chosen reference state and the adjacency
matrix of the graph and calculate the Stieltjes function (for more
details see \cite{js2}). For instance, we consider the finite path
graph and choose the second vertex of the graph as the starting site
of the walk, then the graph does not satisfy three term recursion
relations, i.e., the adjacency matrix has not tridiagonal form.

In Ref. \cite{js2}, it has be shown that the QD parameters for
$\textsf{P}_n$ are given by $\alpha_i=0$ for $i=0,1,...,2k-1$ and\\
$\omega_{2i}=\frac{i}{i+1}, \;\;\ \omega_{2i-1}=\frac{i+1}{i},\;\;\ \omega_{2k-1}=\frac{1}{k}; \;\;\ i=1,...,k-1.$\\
for even values of $n$, where for odd values of $n$ we have \\$\omega_{2i}=\frac{i}{i+1},\;\;\;\ i=1,..., k-1.\\
\omega_{2i-1}=\frac{i+1}{i},\;\;\;\;\ i=1,...,k$,\\
 Substituting these QD parameters in (\ref{op}) and (\ref{oq})
and using (\ref{sti}), the Stieltjes function is obtained as
\begin{equation}
G_\mu(x)=\frac{xU_{n-2}(x/2)}{U_n(x/2)}
\end{equation}
where, $U_n$'s are Tchebichef polynomials of the second kind.
Therefore, by using (\ref{amp0}) we have
\begin{equation}
\hat{q}_0(s)=\frac{-sU_{n-2}(is/2)}{U_n(is/2)}.
\end{equation}
Then, the probability amplitude of the walk at starting site (the
second vertex) at time $t$ is given by
\begin{equation}
q_0(t)
=\frac{1}{n+1}\sum_{l=1}^{n}\sin^2(\frac{2l\pi}{n+1})e^{-2it\cos{l\pi/(n+1)}},
\end{equation}
again one can calculate the other probability amplitudes by using
the Eq.(\ref{Qll}).
\section{Conclusion}
CTQW on graphs was investigated by using the techniques based on
spectral analysis of the graph and inverse Laplace transformation
of the Stieltjes function associated with the graph. It was shown
that the probability amplitudes of observing the walk at a given
site at time $t$ can be evaluated only by calculating the inverse
Laplace transformation of the function $iG_{\mu}(is)$ (without any
knowledge about the spectrum of the graph), where $G_{\mu}(x)$ is
the Stieltjes function associated with the
graph.\\\\

 \vspace{0.5cm}\setcounter{section}{0}
 \setcounter{equation}{0}
 \renewcommand{\theequation}{A-\roman{equation}}
  {\Large{Appendix A}}\\
In this appendix, we give the Laplace transform of the probability
amplitude of observing the walk at starting site at time $t$,
$\hat{q}_0(s)$, for some important
distance-regular graphs with $n\leq 70$.\\\\

\tiny{
\begin{tabular}{|c|c|c|c|c|c|}
 \hline
  \tiny{The graph with Ref.}&\tiny{Intersection  array} & $\hat{q}_0(s)$ & $q_0(t)$ \\
  \hline
  Icosahedron\cite{14}&$\small{\{5,2,1;1,2,5\}}$&$\frac{-i(is^3-4s^2+5is-10)}{s^4+4is^3+10s^2+20is+25}$&$\small{\frac{1}{12}(5e^{it}+e^{-5it}+6\cos\sqrt{5}t)}$ \\
  L(Petersen)\cite{14} &$\small{\{4,2,1;1,1,4\}}$&$\frac{-i(is^3-3s^2+4is-4)}{s^4+3is^3+8s^2+12is+16}$ & $\frac{1}{15}(4e^{it}+e^{-4it}+10\cos2t)$ \\
  Pappus,$3$-cover$K_{3,3}$\cite{14}&$\small{\{3,2,2,1;1,1,2,3\}}$&$\frac{s^4+9s^2+6}{s(s^4+12s^2+27)}$&$\frac{1}{18}(\cos3t+\cos\sqrt{3}t+2)$ \\
  $IG(AG(2,4)\setminus pc)\cite{7}$ & $\{4,3,3,1;1,1,3,4\}$& $\frac{s^4+16s^2+12}{s(s^4+20s^2+64)}$ & $\frac{1}{16}(\cos4t+12\cos2t+3)$ \\
  $3$-cover$K_{9,9}$\cite{vandam}& $\{9,8,6,1;1,3,8,9\}$& $\frac{s^4+81s^2+216}{s(s^4+90s^2+729)}$ & $\frac{1}{27}(\cos9t+18\cos3t+8)$ \\
  Odd($4$)\cite{3}& $\{4,2,1;1,1,4\}$ & $\frac{i(s^4+3is^3+13s^2+15is+20)}{is^5-3s^4+18is^3-30s^2+65is-75}$ & $\frac{1}{15}(4e^{it}+e^{-4it}+10\cos2t)$ \\
  $SRG\setminus$spread\cite{10}& $\{9,6,1;1,2,9\}$& $\frac{-i(is^3-8s^2+9is-18)}{s^4+8is^3+18s^2+72is+8}$ & $\frac{1}{40}(9e^{it}+e^{-9it}+30\cos3t)$ \\
  $3$-cover$K_{6,6}$\cite{vandam}& $\{6,5,4,1;1,2,5,6\}$&$\frac{-i(is^3-16s^2+17is-136)}{s^4+16is^3+34s^2+272is+289}$& $\frac{1}{36}(2\cos6t+24\cos\sqrt{6}t+10)$\\
  Hadamard graph\cite{vandam} & $\{12,11,6,1;1,6,11,12\}$& $\frac{s^4+144s^2+792}{s(s^4+156s^2+1728)}$ & $\frac{1}{24}(\cos12t+12\cos2\sqrt{5}t+8)$ \\
  $IG(AG(2,5)\setminus pc)$\cite{7} & $\{5,4,4,1;1,1,4,5\}$& $\frac{s^4+25s^2+20}{s(s^4+30s^2+125)}$ & $\frac{1}{25}(\cos5t+20\cos\sqrt{3}t+11)$ \\
  Hadamard graph\cite{9} & $\{8,7,4,1;1,4,7,8\}$& $\frac{s^4+64s^2+224}{s(s^4+72s^2+512)}$ & $\frac{1}{32}(2\cos8t+16\cos2\sqrt{2}t+14)$ \\
  Desargues\cite{14} & $\small{\{3,2,2,1,1;1,1,2,2,3\}}$ & $\frac{s(s^4+11s^2+22)}{s^6+14s^4+49s^2+36}$ & $\frac{1}{10}(\cos3t+4\cos2t+10\cos t)$ \\
  Klein\cite{14} & $\{7,4,1;1,2,7\}$& $\frac{-i(is^3-6s^2+7is-14)}{s^4+6is^3+14s^2+42is+49}$ & $\frac{1}{24}(7e^{it}+e^{-7it}+16\cos\sqrt{7}t)$ \\
  $H(3,3)$\cite{14}& $\{6,4,2;1,2,3\}$& $\frac{-i(is^3-6s^2+3is-24)}{s(s^3+6is^2+9s+54i)}$ & $\frac{1}{27}(e^{-6it}+8e^{3it}+6e^{-3it}+12)$ \\
  coxeter\cite{14} & $\{3,2,2,1;1,1,1,2\}$& $\frac{i(s^4+2is^3+5s^2+6is+2)}{is^5-2s^4+8is^3-12s^2+11is-6}$ & $\frac{1}{28}(19e^{it}+8e^{-2it}+12\cos\sqrt{2}t)$ \\
  Mathon(Cycl$(13,3)$)\cite{17} & $\{13,8,1;1,4,13\}$& $\frac{-i(is^3-12s^2+13is-52)}{s^4+12is^3+26s^2+156is+169}$ & $\frac{1}{42}(13e^{it}+e^{-13it}+28\cos\sqrt{13}t)$ \\
  Taylor($P(17)$)\cite{vandam} & $\{17,8,1;1,8,17\}$& $\frac{-i(is^3-3s^2+4is-4)}{s^4+3is^3+8s^2+12is+16}$ & $\tiny{\frac{1}{36}(17e^{it}+e^{-17it}+18\cos\sqrt{17}t)}$ \\
  Taylor(SRG$(25,12)$)\cite{vandam}& $\{25,12,1;1,12,25\}$& $\frac{-i(is^3-24s^2+25is-300)}{s^4+24is^3+50s^2+600is+625}$ & $\frac{1}{52}(25e^{it}+e^{-25it}+26\cos5t)$ \\
  Mathon(Cycl$(16,3)$)\cite{17}&$\{16,10,1;1,5,16\}$&$\frac{-i(is^3-15s^2+16is-80)}{s^4+15is^3+32s^2+240is+256}$ & $\frac{1}{51}(16e^{it}+e^{-16it}+34\cos4t)$ \\
  Mathon(Cycl$(11,5)$)\cite{17}& $\{11,8,1;1,2,11\}$& $\frac{-i(is^3-10s^2+11is-22)}{s^4+10is^3+22s^2+110is+121}$ & $\frac{1}{60}(11e^{it}+e^{-11it}+48\cos\sqrt{11}t)$ \\
  Mathon(Cycl$(19,3)$)\cite{17}& $\{19,12,1;1,6,19\}$& $\frac{-i(is^3-18s^2+19is-114)}{s^4+18is^3+38s^2+342is+361}$ & $\frac{1}{60}(19e^{it}+e^{-19it}+40\cos\sqrt{19}t)$ \\
  Taylor($SRG(29,14)$)\cite{vandam}& $\{29,14,1;1,14,29\}$& $\frac{-i(is^3-28s^2+29is-406)}{s^4+28is^3+58s^2+812is+841}$ & $\frac{1}{60}(29e^{it}+e^{-29it}+30\cos\sqrt{29}t)$ \\
  Taylor($P(13)$)\cite{vandam} & $\{13,6,1;1,6,13\}$& $\frac{-i(is^3-12s^2+13is-78)}{s^4+12is^3+26s^2+156is+169}$&$\tiny{\frac{1}{28}(13e^{it}+e^{-13it}+14\cos\sqrt{13}t)}$ \\
  $GQ(2,4)\setminus$spread\cite{14}& $\{8,6,1;1,3,8\}$& $\frac{-i(is^3-5s^2+22is-8)}{s^4+5is^3+30s^2+40is+64}$ & $\tiny{\frac{1}{27}(8e^{it}+e^{-8it}+12e^{-2it}+6e^{4it})}$ \\
  Doro &$\{12,10,3;1,3,8\}$& $\frac{-i(is^3-11s^2+20is-120)}{s(s^3+11is^2+32s+240i)}$ & $\frac{1}{68}(e^{-12it}+17e^{-4it}+16e^{5it}+34)$ \\
  Locally Petersen&$\{10,6,4;1,2,5\}$ &  $\frac{-i(is^3-12s^2-15is-60)}{s(s^3+12is^2-5s+150i)}$ & $\frac{1}{65}(e^{-10it}+13e^{-5it}+25e^{3it}+26)$ \\
  Taylor(GQ$(2,2)$)\cite{vandam} & $\{15,8,1;1,8,15\}$& $-\frac{is^3-12s62+43is-90}{s^4+12is^3+58s^2+180is+225}$ & $\tiny{\frac{1}{32}(15e^{it}+6e^{5it}+10e^{-3it}+e^{-15it})}$ \\
  Taylor($T(6)$)\cite{vandam} & $\{15,6,1;1,6,15\}$& $\frac{-i(is^3-16s^2-13is-120)}{s^4+16is^3+2s^2+240is+225}$ & $\tiny{\frac{1}{32}(15e^{it}+10e^{3it}+6e^{-5it}+e^{-15it})}$ \\
  Gosset,Tayl(Schl$\ddot{a}$fli)\cite{vandam}& $\{27,10,1;1,10,27\}$& $\frac{-i(is^3-32s^2-129is-432)}{s^4+32is^3-102s^2+864is+729}$ &$\frac{1}{56}(27e^{it}+e^{-27it}+7e^{-9it}+21e^{3it})$ \\
  Taylor(Co-Schl$\ddot{a}$fli)\cite{vandam}& $\{27,16,1;1,16,27\}$& $\frac{-i(is^3-20s^2+183is-270)}{s^4+20is^3+210s^2+540is+729}$ & $\frac{1}{56}(27e^{it}+e^{-27it}+7e^{9it}+21e^{-3it})$ \\
  $GH(2,2)$\cite{vandam} &$\{6,4,4;1,1,3\}$& $\frac{-i(is^3-5s^2+9is-21)}{s^4+5is^3+15s^2+45is+54}$ & $\frac{1}{63}(27e^{it}+e^{-6it}+14e^{3it}+21e^{-3it})$ \\
  $H(3,4)$,Doob\cite{vandam}&$\{9,6,3;1,2,3\}$& $\frac{-i(is^3-12s^2-23is-42)}{s^4+12is^3-14s^2+132is-135}$ & $\frac{1}{64}(27e^{-it}+27e^{3it}+9e^{-5it}+e^{-9it})$ \\
  Wells\cite{7}& $\{5,4,1,1;1,1,4,5\}$& $\frac{i(s^4+3is^3+13s^2+15is+20)}{is^5-3s^4+18is^3-30s^2+65is-75}$ &$\tiny{\frac{1}{32}(10e^{-it}+e^{-5it}+5e^{3it}+16\cos\sqrt{5}t)}$ \\
  $GH(2,1)$\cite{14}& $\small{\{4,2,2;1,1,2\}}$& $\frac{-i(is^3-4s^2+is-6)}{s^4+4is^3+5s^2+18is+8}$ & $\tiny{\frac{1}{21}(e^{-4it}+8e^{2it}+12(e^{-it}+\cos\sqrt{2}t))}$ \\
  $GH(3,1)$\cite{13}& $\{6,3,3;1,1,2\}$ & $\frac{-i(is^3-8s^2-11is-8)}{s^4+8is^3-5s^2+44is-12}$ & $\frac{1}{52}(e^{-6it}+27e^{2it}+24e^{-2it}+24\cos\sqrt{3}t)$ \\
  Dodecahedron\cite{14} & $\small{\{3,2,1,1,1;1,1,1,2,3\}}$& $\frac{-i(is^5-2s^4+7is^3-19s^2+10is-6)}{s(s^5+2is^4+10s^3+16is^2+25s+30i)}$ & $\tiny{\frac{1}{20}(5e^{-it}+4e^{2it}+e^{-3it}+6\cos\sqrt{5}t+4)}$ \\
  Perkel\cite{5}& $\{6,5,2;1,1,3\}$& $\frac{-i(is^3-6s^2+2is-15)}{s^4+8s^2+6is^3+51is-18}$ & $\frac{1}{57}(e^{-6it}+20e^{3it}+36e^{-3it/2}+36\cos\sqrt{5}/2t)$ \\
  $GO(2,1)$\cite{7} & $\{4,2,2,2;1,1,1,2\}$& $\frac{i(s^4+5is^3-s^2+13is-2)}{is^5-5s^4+3is^3-29s^2+2is-24}$ & $\frac{1}{45}(9e^{it}+10e^{-it}+16e^{2it}+9e^{-3it}+e^{-4it})$ \\
  $3$-cover$GQ(2,2)$\cite{7}&$\{6,4,2,1;1,1,4,6\}$& $\frac{i(s^4+5is^3+11s^2+33is+6)}{is^5-5s^4+17is^3-57s^2+72is-108}$ & $\frac{1}{45}(9e^{-it}+18e^{2it}+5e^{3it}+12e^{-3it}+e^{-6it})$ \\
  $J(8,4)$\cite{vandam}&$\{16,9,4,1;1,4,9,16\}$& $\frac{i(s^4+20is^3-44s^2+368is-192)}{is^5-20s^4-28is^3-592s^2-128is-2048}$ & $\frac{1}{70}(e^{-16it}+7e^{-8it}+28e^{2it}+20e^{-2it}+14e^{4it})$ \\
  \hline
\end{tabular}}\\

\vspace{1cm} \setcounter{section}{0}
 \setcounter{equation}{0}
 \renewcommand{\theequation}{A-\roman{equation}}
 \small{
}
\end{document}